\documentclass[usegraphicx]{mn2e}
\usepackage{graphicx}
 
\title{Bulk motion Comptonization in black-hole accretion flows}

\author[A.~Nied\'zwiecki and A.~A.~Zdziarski]
{Andrzej Nied\'zwiecki$^1$\thanks{E-mail: amn@kfd2.fic.uni.lodz.pl, aaz@camk.edu.pl} 
and  Andrzej A.~Zdziarski$^2$\footnotemark[1] \\
$^1$\L \'od\'z University, Department of Physics, 
Pomorska 149/153, 90-236 \L \'od\'z, Poland\\
$^2$Centrum Astronomiczne im.\ M. Kopernika, Bartycka 18, 00-716 Warszawa, Poland
}

\date{Accepted 2005 October 18. Received 2005 October 11; in original form 2005 July 22}

\pagerange{\pageref{firstpage}--\pageref{lastpage}}
\pubyear{2005}
\begin{document}
\maketitle
\label{firstpage}

\begin{abstract}
We study spectra generated by Comptonization of soft photons by cold
electrons radially free-falling onto a black hole. We use a Monte
Carlo method involving a fully relativistic description of
Comptonization in the Kerr space-time. In agreement with previous
studies, we find that Comptonization on the bulk motion of free fall
gives rise to power-law spectra with the photon index of $\Gamma\ga
3$.  In contrast to some previous studies, we find that these
power-law spectra extend only to energies $\ll m_{\rm e} c^2$. We
indicate several effects resulting in generic cutoffs of such spectra
at several tens of keV, regardless of any specific values of physical
parameters in the model. This inefficiency of producing photons with
energies $\ga 100$ keV rules out bulk motion Comptonization as a main
radiative process in soft spectral states of black-hole binaries. The
normalization of the power law (below the cutoff) with respect to the
peak of the blackbody emission of the surrounding disc is typically
very low, except for models with an overlap between the disc and the
plasma, in which case the spectra are very soft, $\Gamma\ga 4$.
\end{abstract}

\section{Introduction}
\label{intro}

Bulk motion Comptonization (BMC) was first considered in a series of
papers by  Blandford \& Payne (1981a, b) and Payne \& Blandford
(1981).  They derived the transfer equation in the diffusion
approximation for photons repeatedly upscattered by cold electrons
undergoing converging inflow. Refining that approach, Psaltis \& Lamb
(1997) introduced relativistic corrections to that transfer equation,
solved by Psaltis (2001) in a flat spacetime. Solution of the transfer
equation yields a power-law spectrum with a photon spectral index of
$\Gamma=3$ for the velocity profile corresponding to free
fall. Titarchuk \& Zannias (1998) and Turolla, Zane \& Titarchuk
(2002) showed that this property remains valid when effects of general
relativity are taken into account. While the above studies demonstrate
the ability of this model for producing power-law spectra by
scattering on the bulk motion, they do not determine how far such a
spectrum extends to high energies. In particular, the Compton recoil
and other effects resulting in high energy turnover of the spectrum
were not taken into account in those papers.

Notwithstanding the uncertainty about that issue, Chakrabarti \&
Titarchuk (1995) and  Ebisawa, Titarchuk \& Chakrabarti (1996)
proposed the BMC to be responsible for producing high-energy tails in
the soft spectral states of black-hole binaries. They argued that both
Keplerian and sub-Keplerian (close to free fall) components are always
present in black-hole accretion flows. In the hard state, the latter
is hot, and thermal Comptonization dominates.  In the soft state, the
sub-Keplerian flow remains cold and Comptonization on the bulk motion
explains the presence of high-energy tails, which are observed to have
typically $\Gamma\simeq 2.5$--3.5. Then, application of their BMC
model to observational data of several sources was presented by, e.g,
Shrader \& Titarchuk (1998, 1999) and Borozdin et al.\ (1999).
 
On the other hand, e.g., Zdziarski (2000), Zdziarski et al.\ (2001),
McConnell et al.\ (2002) and Zdziarski \& Gierli\'nski (2004) pointed
out that the high-energy tails observed in the soft states of black
hole binaries extend to energies well above $\sim$0.5 MeV (see also
Grove et al.\ 1998; Tomsick et al.\ 1999; Ueda et al.\ 2002), without
showing any signature of a high-energy cutoff. In particular, the
high-energy tail of the black-hole binary Cyg X-1 in the soft state
shows no cutoff up to 10 MeV (McConnell et al.\ 2002).

Another disagreements between theoretical predictions of the BMC model
and observational data, noted by Zdziarski (2000), concerns Compton
reflection (e.g., Magdziarz \& Zdziarski 1995) which is strong in the
soft state of black-hole binaries (e.g., Gierli\'nski et al.\ 1999),
while it should be close to null according to the BMC
model. Furthermore, the presence of the tail in the soft state was
claimed to be a unique black hole signature as the BMC process
requires the presence of a horizon (e.g., Laurent \& Titarchuk 1999,
hereafter LT99). However, by now, observations of weakly-magnetized
accreting neutron-star binaries in the soft state show high energy
tails to be common (e.g., Di Salvo et al.\ 2000, 2001, 2002; D'Amico
et al.\ 2001; Iaria et al.\ 2001; Farinelli et al.\ 2005).

Still, the cutoff energy seems to be a key feature allowing to test
the model against data. However, its determination has remained
relatively uncertain. The studies dealing with the position of the
high energy cutoff include an approximate analytic analysis in
Titarchuk, Mastichiadis \& Kylafis (1997) and Monte Carlo simulations
by  LT99 and Laurent \& Titarchuk (2001). The Monte Carlo calculations
of LT99 have been then compared with the {\it CGRO}/OSSE soft-state
spectra of the black-hole binaries GRO J1655--40 and GRS 1915+105 by
Zdziarski (2000) and Zdziarski et al.\ (2001), respectively, showing
that the BMC predicted spectra fall below of the observed ones at
energies $\ga$100 keV.

Titarchuk et al.\ (1997) have presented a simple formula  for the
cutoff energy derived under the assumption that the cutoff occurs at
the energy for which energy gains are balanced by energy losses in the
plasma rest frame, but ignoring light bending as well as photon
trapping.  We find that their formula significantly overestimates the
cutoff energy even if these effects are not important.  LT99 use a
fully relativistic description of photon propagation and scattering in
the Schwarzschild metric. However, their model appears to oversimplify
the treatment of photon transfer in an accreting plasma, as we discuss
in Section \ref{models}.

In this paper, we study formation of BMC spectra using a Monte Carlo
method  described in Nied\'zwiecki (2005, hereafter N05) in a model
with radially  free-falling electrons in a plasma around a black
hole. The plasma Comptonizes  seed photons from an optically-thick
disc.  Our model involves a fully general  relativistic (GR)
description of photon transfer and Compton scatterings in a  plasma
located close to a Kerr black hole. In addition, for illustrative
purposes, we calculate spectra from a model with photon trajectories
approximated by straight lines.

\section{The Models}
\label{models}

We consider a black hole accreting matter at a mass accretion rate,
$\dot M$. The black hole is characterized by its mass, $M$, and
angular momentum, $J$. In our model we use three reference frames.
Photon trajectories are described in the Boyer-Lindquist coordinate
system, $x^i = (t,R,\theta,\phi)$  (which generalizes the
Schwarzschild coordinate system for $J \neq 0$). Physical processes
(in particular Compton scattering) are described in local rest frames
of the matter. Finally, reference frames of the  locally non-rotating
observers (observers with constant $r$ and $\theta$ but dragged in the
azimuthal direction; in the Schwarzschild metric they are equivalent
to static observers, i.e., observers with $r,\theta,\phi$=const),
introduced by Bardeen, Press \& Teukolsky (1972), are auxiliary in our
model.  We use the following dimensionless parameters:
\begin{equation}
r = {R \over R_{\rm g}},~~~\hat t = {ct \over R_{\rm g}},
~~~{\dot m} = { {\dot M} \over {\dot M}_{\rm E}},
~~~a = {J \over c R_{\rm g} M},
\end{equation}
where ${\dot M}_{\rm E} = 4 \pi G M m_{\rm p} / (\sigma_{\rm T} c)$ is
the Eddington accretion rate and $R_{\rm g} = GM/c^2$ is the
gravitational radius.  The formalism used by us does not work for
$a=0$ (specifically, the turning point in $\theta$-motion is found
according to the formalism described in Chandrasekhar 1983, which
requires $a \neq 0$); therefore, spectra for a non-rotating black hole
are obtained assuming $a=10^{-3}$.

For most of spectra presented in this paper, we assume that a
Keplerian,  optically thick, disc is replaced by a free-falling plasma
within $r_{\rm tr}$.  We assume $r_{\rm tr}=20$, for which a
relatively high fraction of disc photons is scattered by the
plasma. The plasma is assumed to be spherical, neglecting its likely
flattening. Both these assumptions maximize the importance of the BMC
for spectral formation.  In Section \ref{illumination}, we consider
also a model where the disc (surrounded by the free-falling plasma)
extends down to the event horizon.

Velocity field of the free-fall, with null angular momentum, is given
by
\begin{equation}
u^t     = {A \over \Sigma \Delta},~~
u^r     = - {(A - \Delta \Sigma)^{1/2} \over  \Sigma},~~
u^{\theta} = 0,~~
u^{\phi}= {2 a r \over \Sigma \Delta},
\end{equation}
where 
\[
\Delta=r^2 - 2r +a^2,~~~\Sigma= r^2 + a^2 \cos^2 \theta,
\]
\begin{equation}
A=(r^2 + a^2)^2 - a^2 \Delta \sin^2 \theta, 
\label{delsig}
\end{equation}
yielding the velocity in the locally non-rotating frame, 
\begin{equation}
\beta^r \equiv {v^r \over c} 
= {A^{1/2} \over \Delta} {u^r \over u^t},~~ v^{\theta} = 0,~~ v^{\phi} = 0,
\label{v}
\end{equation}
with the corresponding Lorentz factor, $\gamma = \left[ 1 - (\beta^r)^2 \right] ^{-1/2}$.

We assume that the rest density of electrons, $n$, is uniform on
surfaces of constant $r$, and determine the density radial profile
from the continuity equation.  Assuming a pure H plasma, where the
comoving mass density is $\rho = n m_{\rm p}$, and integrating the
mass conservation equation, we obtain,
\begin{equation}
n(r) = { 2 {\dot m} \over R_{\rm g} \sigma_{\rm T} \int_0^\pi 
|g|^{1/2}|u^r|  {\rm d} \theta},
\label{n0}
\end{equation}
where $g$  is the determinant of the metric, $|g| = \Sigma^2 \sin^2
\theta$. Below we also use a dimensionless density parameter,
\begin{equation}
\hat n \equiv n R_{\rm g} \sigma_{\rm T}.
\label{nhat}
\end{equation}
For $a=0$, equations (\ref{v}) and (\ref{n0}) yield $\beta^r =
-(2/r)^{1/2}$ and $\hat n = (2/r)^{1/2} \dot m/2r$, respectively, and
for $a=0.998$, departures from these profiles are negligible for 
$r > 3$.

We assume that the electron temperature, $T_{\rm e}$, is constant in
the inner flow. For most of spectra presented in this paper, we assume
$kT_{\rm e}=5$ keV, at which the BMC dominates over thermal effects
(e.g., Blandford \& Payne 1981a; Ebisawa et al.\ 1996).

In all simulations, seed photons come from blackbody emission of the
Keplerian disc. Except for a model with the free-falling plasma
overlapping with the disc (Section \ref{illumination}),  we take into
account emission from the area of the disc between $r_{\rm tr}$ and
$r_{\rm out} = 100$. The latter value is chosen because illumination
of the inner  cloud by emission from $r > 100$ is negligible and thus
an increase of $r_{\rm out}$ above 100 does not affect resulting BMC
spectra.  The thermal emission of the disc is modelled as in N05;
namely, the point of emission of each photon is generated according to
the radial distribution of the disc emissivity (Page \& Thorne 1974)
and the photon energy (in the disc local rest frame) is generated from
the blackbody distribution corresponding to the local temperature.
Note that the radial emissivity (and hence the temperature profile) is
derived with the boundary condition of the vanishing stress at $r_{\rm
ms}$ (where $r_{\rm ms}$  is the radius of the marginally stable
orbit), valid for a disc extending down to $\leq r_{\rm ms}$.  In
models with the disc truncated at $r_{\rm tr} > r_{\rm ms}$,
deviations from this radial dependence may occur, e.g., a decrease of
the energy emitted by the disc due to a lower amount of mechanical
energy transported outward. On the other hand, magnetic stresses may
be important near the $r_{\rm ms}$, modifying somewhat the above
boundary condition and thus the emissivity profile (e.g., Merloni \&
Fabian 2003 and references therein). However, quantitative description
of these effects would require hydrodynamical models of the disc and
its transition into the free-falling plasma, which is beyond the scope
of this paper.

Photons illuminating the region of the free-fall are traced through
their consecutive scatterings until they are captured by the black
hole, hit the disc, or escape the flow. Compton reflection and
reprocessing by the disc are not taken into account. The spectra
presented here correspond to the escaping photons as seen by a distant
observer and are angle-averaged. Simulation of Compton scattering is
performed, in the plasma rest frame, according to the procedure
described in G\'orecki \& Wilczewski (1984). In modelling
Comptonization spectra, we consider two cases, one with photon
trajectories approximated by straight lines, and the other with a full
GR description of photon motion, hereafter referred to as the flat and
GR models, respectively.

Our GR model fully follows the procedure  described in  N05, involving
solution of equations of photon motion in a curved space-time.  The
optical depth along a photon trajectory in a radially inflowing plasma
is given by,
\begin{equation} 
   {\rm d} \tau =   \hat n {\sigma(E_{\rm rest}) \over \sigma_{\rm T}}
     \gamma \left( {\Delta \Sigma \over A} \right)^{1/2} 
    \left( 
    {{\rm d} \hat t \over {\rm d} \zeta}  - 
{A^{1/2} \beta^r \over \Delta}{{\rm d} r \over {\rm d} \zeta} 
              \right) 
      {\rm d} \zeta.
\label{tau}
\end{equation}
This is derived, similarly as in N05, from the elementary expression
for the probability of scattering in the plasma rest frame,  ${\rm d}
\tau = n \sigma {\rm d} l$, where  ${\rm d} l$ is the length of photon
trajectory measured in the plasma  rest frame. In the above equation,
$\zeta$ is an affine parameter, $\sigma$ is the Klein-Nishina
cross-section averaged over the Maxwellian distribution of electron
velocities [see eq.\ (8) in G\'orecki \& Wilczewski (1984)] and
$E_{\rm rest}$ is photon energy in the plasma rest frame (the energy
is affected by both the gravitational  shift and Doppler shift related
to the free-fall velocity).  Note that increase of the optical depth
in equation (\ref{tau}) is written in terms of the increase of
Boyer-Lindquist coordinates (which describe motion 
relative to a distant observer) along the photon trajectory; the
derivatives ${\rm d} r / {\rm d} \zeta$ and  ${\rm d} \hat t / {\rm d}
\zeta$ are determined by equations of  motion, see eq.\ (8) in N05.

On the other hand, photon trajectories in the flat model are
approximated by straight lines and the optical depth has the form
obtained from equation (\ref{tau}) for $r \gg 1$,
\begin{equation}
{\rm d} \tau = n  \sigma(E_{\rm rest})  \gamma \left( 1 - \beta^r \cos
\phi_r  \right) {\rm d}l',
\label{tau_flat}
\end{equation}
where  $\phi_r$ is the angle between the outward radial direction and
the direction of  photon motion in the static frame and ${\rm d}l'$ is
the distance measured  by a static observer (in the flat model,
reference frames of static and distant observers are equivalent).
Note that in our derivation of ${\rm d} \tau$, basing on the
probability of interaction in the plasma rest frame,  the kinematic
term, $\gamma ( 1 - \beta^r \cos \phi_r)$ (e.g., Blumenthal \& Gould
1971; Weaver 1976), results from transformation of the length of a
photon path between the plasma rest frame and the frame of a static
observer. This term has to be taken into account in Monte Carlo
simulations of relativistic Comptonization (e.g., Pozdnyakov, Sobol'
\& Sunyaev 1983; G\'orecki \& Wilczewski 1984). Equation
(\ref{tau_flat}) can be also derived from the probability of
interaction written in terms of quantities defined in the frame of a
static observer (denoted by an apostrophe), ${\rm d} \tau = \alpha'
{\rm d}l'$, where $\alpha' = \sigma'(E') n'$, taking into account the
Lorentz invariance of $E \alpha$ (e.g.,  Rybicki \& Lightman 1979,
section 4.9).

All the remaining assumptions in the flat model are the same as in the
GR model, in particular, capture of photons crossing the event horizon
(assumed in the flat model at $r_{\rm hor} = 2$ for $a=0$).
Obviously, our flat model is not self-consistent. We consider it in
order to illustrate the crucial impact of space-time curvature on
emerging spectra. Also, a  number of previous studies neglected the
curvature of the space-time and our flat model allows to directly
compare these previous results with our Monte Carlo simulations.

We emphasize that an oversimplified method of integrating the optical
depth along photon trajectories,  neglecting both the gravitational
and kinematic effects, have been used in some studies of BMC.  See,
e.g., eq.\ (2) in LT99 for  the radial Thomson optical depth,  which
follows directly from ${\rm d} \tau = \sigma_{\rm T} n {\rm d} R$,
with $n$ being the rest density and $\sigma_{\rm T}$ giving the
probability  of interaction with electron at rest  but ${\rm d}R$
being the change of the Schwarzschild radial coordinate.

\begin{figure*}
\centerline{ \includegraphics[height=120mm]{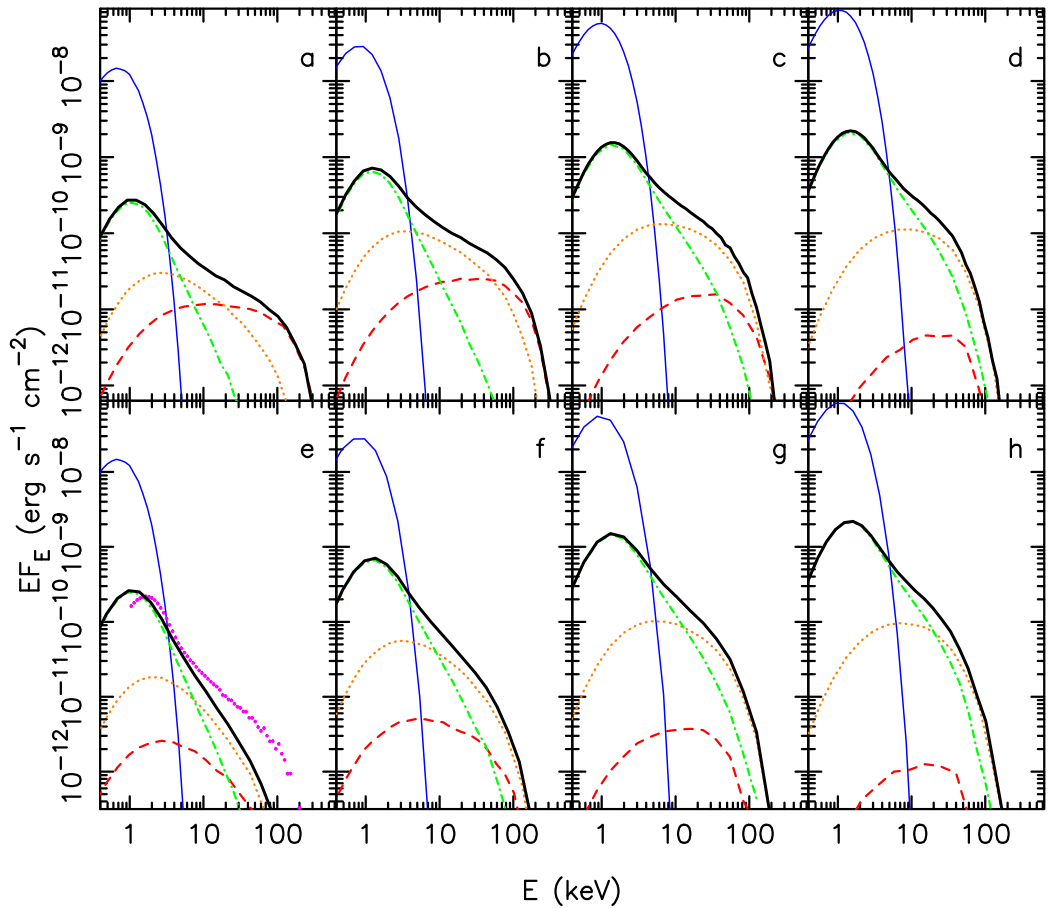} }
\caption{ Emerging spectra for models with a Keplerian disc outside
$r_{\rm tr} = 20$ and a spherical, free-falling, cold ($kT_{\rm e}=5$
keV) inflow inside $r_{\rm tr}$.  The unscattered part of the
blackbody disc emission and the Comptonization spectra are shown by
the thin (blue online) and heavy (black) solid curves, respectively.
All spectra are angle-averaged and correspond to a black hole with  
$M = 10 {\rm M}_{\sun}$ at a distance of $d=5$ kpc. The accretion rate is
$\dot m=2$, 4, 8 and 12 from left to right. The top panels show
spectra for the flat model, and the bottom panels are for the GR model
with a non-rotating black hole.  The dashed (red online), dotted
(orange online) and dot-dashed (green online) curves show contribution
from photons for which the smallest scattering radius is  within
$r=2$--5, 5--8, and 8--20, respectively. The (magenta online) points
in (e) show the spectrum for  $\dot m=2$ and $kT_{\rm e}=5$ keV
obtained by LT99 in their GR model.}
\label{spectra}
\end{figure*}

In order to illustrate the significance of the relativistic terms in
equation (\ref{tau}), we consider now the Thomson optical depth for a
{\it radial\/} trajectory in the Schwarzschild metric. For $a=0$,
$T_{\rm e}=0$ and null angular momentum of a photon, equation
(\ref{tau}) yields
\begin{equation}
{\rm d} \tau = 
\hat n \gamma (1 \pm \beta^r) (1 - 2/r)^{-1/2} {\rm d} r.
\label{radial}
\end{equation}
The last two terms give the proper length in the Schwarzschild
geometry, $g_{rr}^{1/2} {\rm d} r$, where $g_{rr} = (1 -
2/r)^{-1}$, and  $\gamma (1 \pm \beta^r)$ gives the Lorentz
transformation from the plasma rest frame; specifically, $\gamma \hat n$
gives the density of electrons in the frame of a static observer
and $\sigma_{\rm T}(1 \pm \beta^r)$ gives the probability 
of interaction with a beam of electrons moving with
the velocity $\beta^r$ (e.g., Weaver 1976).  For the free-fall
velocity field, $\beta^r = -(2/r)^{1/2}$, and then equation (\ref{radial})
yields
\begin{equation}
{\rm d} \tau = { \hat n {\rm d} r \over 1 \pm (2/r)^{1/2}}.
\label{radial_ff}
\end{equation}
The upper and lower sign in equations (\ref{radial}) and
(\ref{radial_ff}) corresponds to ingoing (${\rm d} r/{\rm d} \zeta<0$) 
and outgoing photons, respectively. (Note that the optical depth
remains finite for ingoing photons at the event horizon.)  
The non-relativistic (for, in particular, $\beta^r\ll 1$) approach, 
${\rm d} \tau = \hat n {\rm d} r$, yields roughly two times higher
optical depth than our equation (\ref{radial_ff}) for ingoing photons,
while for photons escaping from the vicinity of the event horizon, it
underestimates $\tau$ by a factor of a few.  Equation
(\ref{radial_ff}) implies a significant decrease of the probability of
escape from the innermost region of a free-falling flow  with respect
to photon transfer in a static plasma with the same rest density.

Monte Carlo simulations of BMC for both flat and GR (with $a=0$)
models were presented in LT99. In both cases, the assumptions
underlying their simulations are similar as in this paper (they assume
different spatial distribution of the source of soft photons, which,
however, affects only slightly the BMC spectrum, see Section
\ref{illumination}). However, both their flat and GR models appear to
neglect the dependence of the probability of interaction on the angle
between photon momentum and the direction of the bulk motion. As we
discuss above, this may result in  strongly overestimated  emission
from the inner region. In Section \ref{index} below, we indicate this
effect as a possible explanation for discrepancies between results of
our model and LT99 at high accretion rates. On the other hand, the GR
model developed in LT99 takes into account the GR term,
$g_{rr}^{1/2}$, for the proper length in calculation of $\tau$
(L. Titarchuk, personal communication).

In order to thoroughly test our method and results, we have also
developed an independent model for Comptonization in the Schwarzschild
metric (Nied\'zwiecki \& Sitarek, in preparation), which involves a
much simpler description of GR effects than our Kerr metric model. We
have obtained results fully consistent with these  presented here for
$a=10^{-3}$, in particular those where we encountered some
discrepancies with previous studies.

\section{Results}
\label{results}

Spectra emerging in models with $r_{\rm tr} = 20$,  $kT_{\rm e}=5$
keV, $M=10{\rm M}_{\sun}$ and various accretion rates are shown in Fig.\
\ref{spectra}(a--d) for the flat model and in Fig.\ \ref{spectra}(e--h)
for the GR model with a non-rotating  black hole. We also show contributions 
from photons scattered in various parts of the inner flow.
 
\subsection{The cutoff energy in the flat model}
\label{cutoff_flat}

In the flat model, the cutoff energy decreases with increasing $\dot
m$, starting from $\sim$100 keV at low accretion rates to a few tens
of keV at $\dot m \ga 10$.  As illustrated in Fig.\
\ref{spectra}(a--d), this decrease of the cutoff energy  is primarily
related to the diminishing contribution of radiation emerging from
inner parts of the flow.  In the flat model, the innermost parts
generate radiation extending to highest energies.  However,  the
optical depth for photons moving outward in the innermost region is
typically a few times higher than for photons moving inward, see
Section \ref{models}. As a result, photons at high accretion rates are
trapped in the inner flow and carried under the event horizon.  Such
trapping has been considered, e.g., by Begelman (1979) and Payne \&
Blandford (1981), who defined the trapping radius, $r_{\rm trap} \sim
\dot m$,  as the distance at which the diffusion velocity equals the
flow velocity.

An increase of $\dot m$ results then in an increase of the trapping
radius and depletion of increasing number of photons from the
high-energy part of the spectrum.  Moreover, photons that diffusively
escape from the region inside $r_{\rm trap}$ lose a significant
fraction of their energy in scatterings, compare the dashed and the
heavy solid curves in Fig.\ \ref{spectra}(d). In the flat model, the
position of the trapping radius is crucial for the dependence of  the
cutoff energy on the accretion rate.

We note that the formula derived in Titarchuk et al.\ (1997) for BMC
in flat space-time, $E_{\rm cut} =  [4/ \dot m + (4/3)(v^r/c)^2]
m_{\rm e}c^2$  [note  that the form of the second term in their eq.\
(9) is valid only for a plasma with $v^r = -(2/r)^{1/2} c$;  here we
use a more general form which can be derived from Appendix D of their
paper], predicts the cutoff energies a few times higher than these
found in our simulations even when setting $v^r=0$. The cause for that
discrepancy is that their analysis ignores photon trapping and is
based on the assumption that the cutoff occurs at photon energies for
which energy gains are balanced by Compton recoil.  We find that
although the average increase of photon energy in a scattering  indeed
increases with decreasing $\dot m$ (as a lower density results in a
higher average velocity difference between scatterings), the above
formula strongly overestimates the cutoff energy.  Even at low
accretion rates, $\dot m < 4$, for which trapping is not  important,
the formula is not consistent with our results, yielding unrealistic
values of $E_{\rm cut} > 500$ keV, regardless of the value of the
radial velocity.

\begin{figure}
\centerline{\includegraphics[height=65mm]{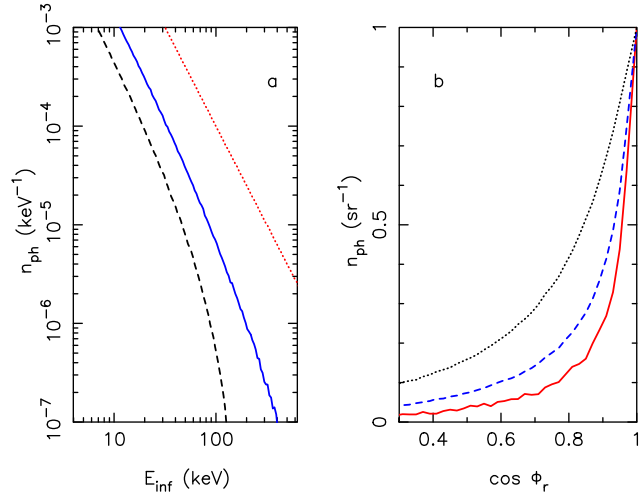}}
\caption{(a) The dashed (black online) and solid (blue online) curves
show the angle-averaged spectra of photons at infinity that scatter at
$3.5 R_{\rm g}$ in the Schwarzschild and extreme Kerr geometry,
respectively, off electrons with the free-fall velocity ($v^r \approx
-0.75c$ in both cases). The incident photons, with the power-law
spectrum shown by the dotted (red online) line, move in the inward
radial direction. (b) The angular distribution of photons in the local
rest frame  scattered within the inner $5 R_{\rm g}$ for $\dot m=4$.
The number of photons per unit solid angle is given as a function of
the angle of the photon direction before scattering with respect to
the radial direction ($\cos \Phi_r = 1$ corresponds to the inward
direction). The dashed (blue online) and solid (red online) curves
show photons with the initial energy corresponding to $E_{\rm inf} <
40$ keV and $>40$ keV, respectively, in the Schwarzschild GR model.
The dotted (black) curve is for photons with $E_{\rm inf} > 40$ keV in
the flat model.  The curves are normalized to unity at $\cos \Phi_r =
1$.}
\label{sch_kerr}
\end{figure}

\subsection{The cutoff energy including space-time curvature}
\label{cutoff_GR}

The spectra in the Schwarzschild GR model are cut off below 100 keV
for all values of the $\dot m$, see Fig.\ \ref{spectra}(e--h). Thus,
bending of photon trajectories strongly diminishes the contribution
from  the innermost region with respect to the flat model. This is due
to several effects.

As explained below, basic properties of photon motion imply that for
$a=0$ all photons escaping the flow from $r\le 5$ (i.e., avoiding
crossing the horizon) are strongly redshifted as seen by a distant
observer. Therefore, the emission from the $r=2$--5 region contributes
negligibly to the overall spectrum in the GR model with $\dot m=2$, in
spite of the trapping surface being very close to the event horizon,
whereas this region gives a major contribution to the 20--200 keV
spectrum in the corresponding flat model, see Figs.\ \ref{spectra}(a), (e).

The observed photon energy, $E_{\rm inf}$,  is related to
the energy in the plasma rest frame, $E_{\rm rest}$, by
\begin{equation}
E_{\rm inf} =  (1-2/r)^{1/2} \gamma (1 - \beta^r \cos \Phi_r) E_{\rm rest},
\end{equation}
where $\Phi_r$ is the angle between the inward radial direction and
the photon direction in the plasma rest frame.  Taking into account
that the Lorentz factor for $\beta^r = -(2/r)^{1/2}$ reduces with the
gravitational redshift, $(1 - 2/r)^{1/2}$ (i.e., their product is unity), 
we obtain,
\begin{equation}
E_{\rm inf} = \left[ 1 + (2/r)^{1/2} \cos \Phi_r \right] E_{\rm rest},
\end{equation}
which implies that only photons emitted inward in the plasma rest
frame ($\Phi_r < \pi/2$) are observed as blueshifted.

On the other hand, the half-angle of the cone of avoidance, as defined in  
Chandrasekhar (1983),
\begin{equation}
\tan \Phi = (r/2 - 1)^{1/2} (r/3 - 1)^{-1} (r/6 + 1)^{-1/2},
\end{equation}
corresponds to $\Phi_r = \pi/2$ [for which the aberration of light
gives $\tan \Phi = 1/(\gamma |\beta^r|)$] at $r=5.2$.  Then, all
photons escaping from $r<5.2$ must be emitted backward and thus they
have $E_{\rm inf}/E_{\rm rest} < 1$. The above constraint is valid for
$a=0$. For $a=0.998$, such an effective redshift of all escaping
photons is restricted only to emission from $r<2$. Note that the
solution of the GR radiative transfer equation by Papathanassiou \&
Psaltis (2001) similarly indicates that, due to the space-time
curvature, the innermost region contributes negligibly to the BMC
spectrum.

Another reason for the inefficiency of BMC is that the free-fall
velocity is relatively small in the region where most of high-energy
emission  is generated. In the GR model, the major contribution to the
high-energy part of the spectrum comes from photons scattered  within
$r=5$--8, where velocity is at most mildly relativistic, $v^r \simeq
0.5$--0.7$c$.

Finally,  we point out the importance of an additional effect,
occurring for photons undergoing scattering  in the Klein-Nishina
regime, which severely reduces emission of hard X-rays.  Namely, the
highest energies are achieved by photons scattered toward the center,
which then must be scattered off the inward direction to escape the
flow.  However, photons are preferentially scattered forward (and
subsequently captured) at high energies.  Then, photons which
significantly change direction in a scattering, so that they can
escape, lose much of their energy to the recoil.  Furthermore, the
decline of the Klein-Nishina cross section makes scattering of
high-energy  photons less probable.

The dashed curve in Fig.\ \ref{sch_kerr}(a) shows the spectrum at
infinity from scattering at $r=3.5$ of radially incoming photons with
a power-law spectrum. At $<$40 keV, the flux of escaping radiation is
over an order of magnitude lower than the illuminating flux due to
capture of the scattered photons.  At $>$40 keV, the preference of
forward scattering combined with the Compton  recoil result in a sharp
spectral cutoff.

Fig.\ \ref{sch_kerr}(b) shows the angular distribution of photons
undergoing scattering  within inner $5 R_{\rm g}$ at $\dot m = 4$.
The incident photons are strongly concentrated along the inward radial
direction, and this concentration increases with the increasing photon
energy.  This radiation anisotropy results in depletion of hard X-rays
from the radiation leaving the innermost region, crucial for the
spectrum.  Note the related steepening above $\sim$40 keV of spectra
shown by dashed curves in Figs.\ \ref{spectra}(e--f).  On the other
hand, the concentration of photons in the radial direction is  less
pronounced in the flat model and thus the above effect is less
important.

\subsection{The BMC luminosity}
\label{luminosity}

In all the models with $r_{\rm tr}=20$, the luminosity of the
Comptonization  component is $<$0.04 of the total luminosity. This
fraction drops to $<$0.01 when the transition occurs at the marginally
stable orbit, $r_{\rm tr}=6$. While the BMC component emerging from a
spherically symmetric inner cloud is roughly isotropic, the flux from
a flat disc observed at an inclination, $i$ (with respect to the
symmetry axis) decreases with increasing $i$ as $F(i) \propto \cos
i$. As a result, the relative flux in the BMC component from systems
observed edge-on is higher, e.g., $\sim$0.1 for $\cos i = 0.1$, than
in the average spectra (presented in this paper).

For $r_{\rm tr} = 20$, $\simeq$0.05 of the photons emitted by the disc
illuminate the free-fall region. The fraction of photons crossing the
free-fall region that are scattered is 40 and 80 per cent for $\dot m
= 2$ and 12, respectively. Also, both the average number of
scatterings and the energy transferred to photons increase with
increasing $\dot m$. E.g., Comptonization increases the energy of
photons by a factor of 1.9 and 4.3 for $\dot m=2$ and 12,
respectively. However, an increase of $\dot m$ also results in an
increase of the trapping radius and a higher fraction of the photons
being captured by the black hole, e.g., 20 and 40 per cent at $\dot
m=2$ and 12, respectively. Moreover, trapping affects primarily
photons scattered at small $r$, which achieve the highest energies and
carry a significant fraction of the total energy transferred to
photons. Again for $\dot m=2$ and 12, 20 and 40 per cent of the
photons that get captured carry 50 and 85 per cent, respectively, of
the total power in scattered photons. The resulting conclusion is that
the luminosity in the BMC component is generally lower than the
(initial) luminosity in the blackbody photons irradiating the plasma.

\begin{figure*}
\centerline{\includegraphics[height=80mm]{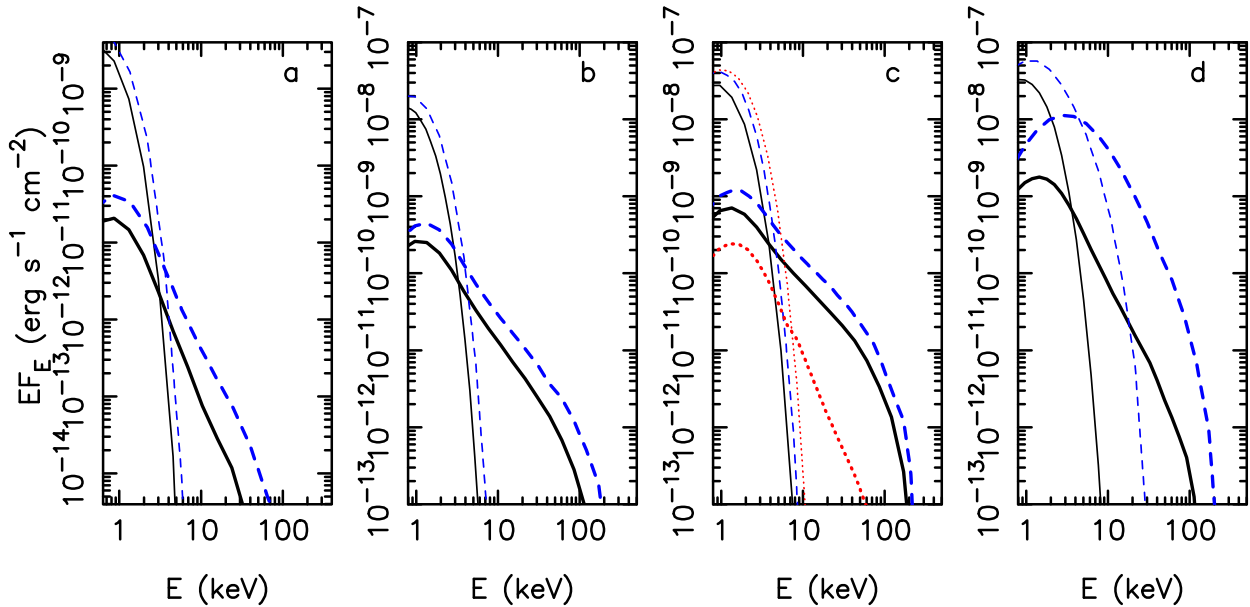}}
\caption{The solid and dashed (blue online) curves show
spectra for the GR model for $a=0$ and $a=0.998$, respectively.  The
thin and heavy curves show the emission from the thermal  disc and
from the BMC, respectively.  The $\dot m=0.5$ (a), 2 (b), 4 (c), and
$r_{\rm tr} = 20$, $M = 10 {\rm M}_{\sun}$,  $kT_{\rm e}=5$ keV.  The
dotted (red online) curves in (c) are for  the GR model with $r_{\rm
tr} = 6$, $\dot m=4$ and $a=0$.  The spectra in (d) are for the model
with the disc extending down to the event horizon described in
Section \ref{illumination}, with $\dot m=4$ and $r_{\rm ff} = 20$.}
\label{sch_kerr_2}
\end{figure*} 

\subsection{Black hole spin}
\label{spin}

Both the radial velocity and density profile changes with varying $a$
are negligible for the considered spherical free-falling plasma. A
significant impact of the space-time metric on BMC spectra arises
only due to properties of photon trajectories in the vicinity of the event
horizon. The number of photons escaping from the innermost region
increases with the increasing $a$ for other parameters unchanged.
Furthermore, the escaping photons have slightly higher blueshifts at
higher values of $a$. For example, in the case of $a=0.998$, the
maximum blueshift, $E_{\rm inf}/E_{\rm rest} = 1.5$, is obtained for
photons escaping from $3<r<8$, while for  $a=0$, the  maximum
blueshift increases from 1 at $r=5.2$ to 1.3 at $r=8$.  Such
differences in photon motion make the steepening of the high-energy
part of the spectrum generated by scattering at $r=3.5$ in the
extreme Kerr metric less pronounced than in the case of the Schwarzschild
metric (Fig.\ \ref{sch_kerr}a).

Fig.\ \ref{sch_kerr_2} compares observed spectra at $a=0$ and
$a=0.998$ for the same accretion rates. In the case of illumination by
the external disc, Fig.\ \ref{sch_kerr_2}(a--c), Comptonization
spectra depend very weakly on $a$. The strongest dependence might have
been expected for low $\dot m$, for which photons scattered close to
the event horizon are not trapped.  For $\dot m = 0.5$, Fig.\
\ref{sch_kerr_2}(a), the spectra in models with $a=0.998$ and $a=0$
are dominated above 15 keV by emission from $r<5$ and $5<r<8$,
respectively.  However, the related difference between these spectra
is rather moderate; most noticeably,  the cutoffs occur at $\sim$30
and $\sim$50 keV for $a=0$ and 0.998, respectively.  Note also that
for such low values of $\dot m$, the BMC process is very inefficient,
yielding spectra with $\Gamma \ga 4$.  For $\dot m \ge 4$, the spectra
are dominated by photons for which the last scattering takes place at
relatively large values of $r$, where the differences between low and
high $a$ are small.  This makes the black hole spin weakly important
for the spectrum.

Strong gravity effects may affect dynamics of Compton scattering in
the ergosphere of a rapidly rotating black hole (Piran \& Shaham
1977).  We find, however, that such effects are not important for the
formation of BMC spectra due to strongly relativistic radial
velocities.  Ergospheric emission  of a cold free-falling plasma is
strongly collimated inward and it does not give any significant
contribution to the observed spectrum.  Thus, we confirm a conclusion
of N05 that a quasi-Keplerian motion is needed  to make radiative
processes in the ergosphere important for spectral formation.

\subsection{Spectral index}
\label{index}

For all spectra, we determine the photon spectral index, $\Gamma$, in
the 10--30 keV energy range. Below 10 keV, contribution from thermal
Comptonization with $kT_{\rm e}=5$ keV is important, and significant
departures from a power-law occur in some models above 30 keV.

The values obtained by us for $\dot m > 4$ in the Schwarzschild GR
model and for $\dot m > 10$ in the flat model agree with $\Gamma = 3$
of the solution of the transfer equation in the diffusion
approximation (e.g., Payne \& Blandford 1981). At low accretion
rates, spectra of the flat model are harder, $\Gamma \simeq 2.6$ for
$\dot m \le 4$, due to the strong contribution from the innermost
region. This appears to be due to neglecting terms $\propto (v/c)^2$
in the transfer equation of Payne \& Blandford (1981), which may
significantly underestimate the change of energy of a photon scattered
in a plasma with a relativistic velocity (Psaltis \& Lamb 1997).

On the other hand, spectra at low accretion rates in the GR model are
much softer, $\Gamma \ga 3.5$ for $\dot m \le 2$. A steepening with
decreasing $\dot m$ also characterizes spectra obtained by LT99 for
their GR model with $kT_{\rm e}=5$ keV (see table 2 in LT99).
However, their spectra are harder than ours at any value of $\dot
m$. At high accretion rates, the difference is moderate, e.g., $\Gamma
= 2.8$ for $\dot m = 7$ in LT99 vs.\ $\Gamma = 3$ in our model, and we
find that it could be accounted for by including the kinematic term in
the scattering probability (see Section \ref{models}).  Indeed, we
obtain $\Gamma = 2.8$ with that term neglected in equation
(\ref{tau}). The same conclusion is valid for the flat model.  Table 1
in LT99 gives $\Gamma=2$ for their flat model with $\dot m = 7$  and
$kT_{\rm e}=5$ keV, while our calculations yield $\Gamma=2.7$ at this
$\dot m$.  However, we obtain similarly hard spectra, $\Gamma = 2.2$
at $\dot m = 4$--8, by neglecting the kinematic term in equation
(\ref{tau_flat}), approximately accounting  for the difference.

At low $\dot m$, the discrepancies between our models and those of
LT99 become more significant, and we have found no explanation for
them. In particular, we obtain $\Gamma = 3.6$ for $\dot m = 2$ whereas
LT99 find $\Gamma = 2.9$. Both of those spectra are compared  on Fig.\
1(e). The (unphysical) modification of ${\rm d}\tau$ described above
does not resolve this discrepancy.

Furthermore, LT99 conclude that the change of the size of the region
containing free-falling plasma between $r_{\rm tr} = 6$ and 40 has a
weak effect on the spectra. We find an opposite property, with the
spectra softening significantly when $r_{\rm tr}$ drops below 10.  For
$r_{\rm tr}$ approaching 6 (at $a=0$), the spectra  soften by at least
$\Delta \Gamma = 1$ with respect to those shown in Fig.\
\ref{spectra}, as illustrated by the lower dotted curve in Fig.\
\ref{sch_kerr_2}(c). The reason for that is obvious; as shown in Fig.\
\ref{spectra}, a major part of the high energy emission is generated
in the flow beyond $6R_{\rm g}$.  Then the Comptonized component
becomes much weaker and softer if the flow shrinks to $r_{\rm tr}=6$.

\subsection{Illumination patterns}
\label{illumination}

We find that spatial distribution of the sources of seed photons on
the surface of the outer disc does not change the shape of the BMC
spectrum.  By replacing emission from the whole outer disc by emission
from its inner edge (at $r_{\rm tr}$), we obtain very similar slopes
and cutoffs of the BMC component.  This is in agreement with previous
studies based on the solution of the radiative transfer equation,
e.g., Titarchuk et al.\ (1997) and Papathanassiou \& Psaltis (2001).

On the other hand, the relative geometry of the disc and the free-fall
plasma may be crucial for the normalization of the BMC emission with
respect to the disc blackbody. The relatively low normalization found
above (as measured by, e.g., the ratio of the $EF_E$ of the BMC
spectrum at the point of intersection with the disc blackbody to the
$EF_E$ at the peak of the disc spectrum) is due to the small solid
angle subtended by the scattering plasma as seen by the outer
disc. This effect was noted, e.g., by Borozdin et al.\ (1999), who
proposed that an overlap between the free-fall plasma with the disc,
resulting in a stronger illumination, is needed to explain certain
observations of soft states of black-hole binaries by the BMC
model. We note that there is indeed some observational evidence for
the disc extending close to the horizon in some black-hole binaries,
e.g., from thermal disc emission and relativistic Fe profiles (Zhang,
Cui \& Chen 1997; Miller et al.\ 2002).

Therefore, we consider a model with a Keplerian, blackbody-emitting,
disc extending down to the minimum stable orbit. Below it, the disc
material continues to flow to the horizon without dissipation (e.g.,
Cunningham 1975; Muchotrzeb-Czerny 1986). We assume that a spherical
plasma surrounds the disc at radii $r\leq r_{\rm ff}$, where a half of
the accreting matter forms the geometrically thin disc and the
remaining half accretes through the spherical free fall. Then, the
density of  the plasma is determined by equation (\ref{n0}) but with a
half of the total $\dot m$.  The disc emits blackbody photons down to
$r_{\rm ms}$ ($=1.23$ and 6 for $a=0.998$ and  0, respectively), but
below $r_{\rm ff}$, the temperature profile corresponds to a half of
the total $\dot m$. The disc absorbs incident photons also below
$r_{\rm ms}$ (as it remains optically thick at the accretion rates we
consider here; Reynolds \& Begelman 1997).

Fig.\ \ref{sch_kerr_2}(d) shows the resulting spectra for a model with
$r_{\rm ff} = 20$ and the (total) $\dot m=4$. The BMC component
contains now 6 and 18 per cent of the total luminosity for $a=0$ and
0.998, respectively. On the other hand, the spectra of this component
are significantly softer (by $\Delta \Gamma \approx  0.5$ for $a=0$)
than in the corresponding (with $\dot m=2$) model with truncated disc.

For $a=0$, $>$70 per cent of the energy dissipated in the disc is
released at $r>20$. Therefore, a disc truncated at $r=20$ has a
similar luminosity as a disc extending down to $r_{\rm ms}$. However,
a significant fraction of soft photons emitted  within the plasma is
scattered in the latter case, which results in the fraction of disc
photons that are Comptonized increasing to 10 per cent (from 3 per
cent for a truncated disc with $\dot m=4$). A related increase of the
peak of the BMC component at $\sim$1.5 keV (formed mostly by the first
scattering order) by a factor of $\sim$3 is clearly seen by comparing
the heavy solid curves in Figs.\ 3(c) and (d).

On the other hand, over 30 per cent of the scattered photons are
absorbed by the part of the disc within the free-fall region.
Moreover, this absorption affects mostly photons scattered in the
inner region. In the corresponding model with a truncated disc, a
significant fraction of photons forming the high-energy part of the
spectrum gets scattered at $r=5$--8 inward but then escape (after
passing the turning point in the $r$-motion), crossing the equatorial
plane. In contrast, those photons are absorbed in the model with the
overlapping disc. On the other hand, this effect is much less
significant for photons at lower energies, typically formed at larger
distances. Thus,  the decreased contribution of high energy photons
results in a softer  spectrum, with $\Gamma=4.1$ vs.\ $\Gamma=3.6$ in
the truncated disc model with the same electron density.

In the case of $a=0.998$, the seed photons are very centrally
concentrated, with over a half of the disc luminosity produced at
$r\le 5$. Thus, they have relatively high energies, of several keV. As
a result, a power-law shape of the spectrum is achieved only in a
narrow energy range of $\sim 20$--70 keV.  Absorption of photons with
highest energies by the disc is significant in this case, as photon
trajectories are strongly bent toward the equatorial plane at small
$r$  in the extreme Kerr metric. A resulting BMC spectrum is shown by
the heavy dashed curve in Fig.\ \ref{sch_kerr_2}(d).

In the overlap model, an increase of $\dot m$  yields a decrease of
the relative normalization of the BMC component for $a=0.998$, e.g.,
its contribution to the total luminosity decreases from 18 to 10 per
cent for $\dot m$ changing from 4 to 24. For $a=0$, this relative
fraction remains approximately constant, at $\sim$6 per cent. The fact
that the efficiency of the process does not increase with increasing
$\dot m$ is due to the increasing number of photons trapped at high
$\dot m$ (Section \ref{luminosity}).

The value of $r_{\rm ff}$ is crucial for the normalization of these
models. A high normalization of the BMC component with respect to the
disc one may be achieved if a very dense, free-falling, plasma extends
out to further distances than considered above. If an optically-thick
free-fall plasma covers the disc surface up to $r_{\rm ff} \ga 100$,
most of disc emission is Comptonized.  We note that the BMC model
makes some specific predictions for such scenario. Namely, the optical
thickness of the free-fall plasma in the region within $r\simeq
20$--100 is a few times lower than the optical thickness of the inner
region, $r\la 20$. Then, if the free-falling plasma  is sufficiently
dense to Comptonize most of disc emission from $r\ga 20$, almost all
of the total emission from the inner region will be trapped.  Then,
any relativistic signals from the innermost region, including
relativistic fluorescent lines or high-frequency QPOs (assuming their
relation to Keplerian frequencies), would not be
observable. Furthermore, the BMC photons observed by a distant
observer would come from scatterings at large $r$ in a plasma with
subrelativistic velocities. Then, spectrum would be precisely
approximated by the flat model, i.e., would have $\Gamma=3$.

LT99 argue that illumination of the central region can significantly
increase if the surrounding disc is geometrically thick.  They
approximate this effect by emission from the inner edge of the disc at
$6R_{\rm g}$. However, emission from the surface of the disc is not
shown in that paper, and thus the actual normalization is uncertain.

Summarizing this Section, we confirm that the relative normalization
increase significantly when the source of seed photons is within the
free-fall region.  On the other hand, we find that then the BMC
spectrum component becomes significantly softer than in corresponding
models without overlap.

\subsection{BMC vs.\ thermal Comptonization}
\label{thermal}

\begin{figure}
\centerline{\includegraphics[height=60mm]{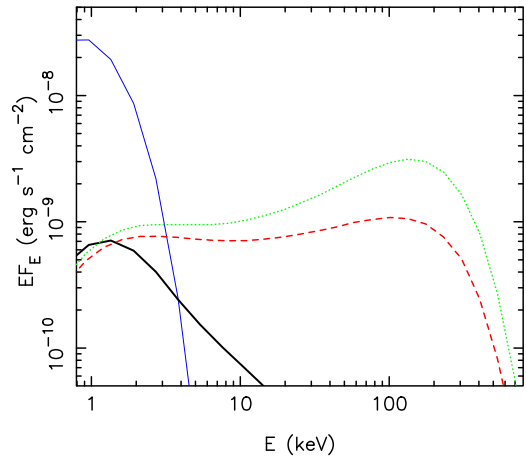}}
\caption{The dotted (green online) and dashed (red online) curves show
Comptonization spectra from static and free-falling, respectively, plasmas
at $kT_{\rm e}=50$ keV. The heavy solid curve
shows the spectrum from a free-falling plasma with $kT_{\rm e}=5$
keV. The free-fall spectra are for $\dot m=4$ and the static plasma
spectrum is for the same density distribution. All models have $a=0$, 
$r_{\rm tr} = 20$ and $M = 10 {\rm M}_{\sun}$. The thin (blue online) solid
curve shows the blackbody disc emission (the same for all
three models).}
\label{comptonization}
\end{figure} 

Fig.\ \ref{comptonization} compares a BMC spectrum with spectra for 
semi-relativistic thermal Comptonization.  For $kT_{\rm e}\ga 10$ keV,
thermal Comptonization completely dominates formation of spectra and
the role of bulk motion is negligible (see also Jaroszy\'nski 2001).

Fig.\ \ref{comptonization} also compares spectra  from thermal
Comptonization between static and  free-falling plasmas with the same
rest density distributions.  The major impact of the free fall on
formation of thermal Comptonization spectra is by trapping 
photons in optically thick regions of inflowing plasma, making a Wien
bump much less pronounced (which effect has been studied by Colpi
1988).

\section{Discussion and Conclusions}

We have examined formation of spectra emerging due to multiple
scattering of  photons in black-hole accretion flows with relativistic
radial velocities. In  agreement with previous studies, we find that
this model gives rise to power-law  spectra with the  photon spectral
index  of $\Gamma \approx 3$ at high $\dot m$, which soften to
$\Gamma>3$ with decreasing $\dot m$. The BMC process cannot  explain
spectra with $\Gamma < 3$. At semi-relativistic temperatures, needed
to obtain such hard spectra by thermal Comptonization, the BMC process
contributes negligibly to the spectral formation.

We point out effects preventing this power-law from extending beyond
$\sim$100 keV. The recoil effect, dominant at lower accretion rates in
flat space-time, results in a cutoff around 100 keV.  At higher
accretion rates, trapping of photons reduces contribution of photons
from innermost regions.  In a curved space-time, bending of photon
trajectories combined with an increased probability of forward
scattering at high energies result in cutoffs at a few tens of keV
even at low accretion rates.  

We do not confirm claims of, e.g., Titarchuk et al.\ (1997) and LT99,
that the power-law spectrum extends to $\sim m_{\rm e} c^2$. 
We find that such high energies are achieved only by photons
scattered toward the center at very small radii, which then do not contribute
to observed spectra.

In models with illumination of the free-falling plasma by the
surrounding disc,  the normalization of the BMC spectral component
with respect to the peak of the disc blackbody spectrum is very low,
typically $\la 10^{-2}$.  If a likely flattening of the plasma were
taken into account, this relative normalization would become even
lower.  Higher normalizations may be obtained in models involving a
disc extending down within the free-falling plasma. In this case, the
high normalization occurs together with soft BMC spectra, $\Gamma \ga
4$, due to absorption of Comptonized photons with highest energies by
the inner disc. The luminosity of the (very soft) BMC component
comparable to the disc one can only be achieved in models with
complete obscuration of the central region, implying then no
observable relativistic signatures.

We find that rotation of the black hole has a minor effect on the
shape BMC spectra. The rotation is only important  for models with the
disc extending close to the event horizon, as in this case
the fraction of soft seed photons emitted in the innermost region increases
with an increase of $a$.

The amplitudes and spectral indices of high energy tails in some
observations of black-hole binaries are consistent with the BMC model
for moderate accretion rates, $\dot m \la 2$. E.g., according to table
1 in Borozdin et al.\ (1999), the 1997 March 24 observation of GRO
J1655--40 is consistent with parameters corresponding to the external
illumination, and observations of 1997 April 24 and 30 are consistent
with  the untruncated disc. However, as discussed in Section
\ref{intro}, determination of the position of the high-energy cutoff
is crucial for the confirmation that those spectra may be explained by
the BMC model. As shown by Tomsick et al.\ (1999), the high-energy
tail of this object extends to $\sim$2 MeV, i.e., much higher than
that possible to obtain in this model. In general, the BMC model is
unable to explain the high energy tails of black-hole binaries in the
soft state, which have often been measured to extend to energies $\ga
500$ keV without any cutoff, and often have too high amplitudes and/or
$\Gamma<3$ (see, e.g., Zdziarski \& Gierli\'nski 2004).

A very low normalization of the high-energy tail (with a poorly
constrained high-energy cutoff) is often seen in the ultrasoft (i.e.,
strongly dominated by a disc blackbody, e.g., Gierli\'nski \& Done
2004) states of black-hole binaries, see fig.\ 8 in Zdziarski \&
Gierli\'nski (2004). However, those tails appear rather hard, e.g.,
with $\Gamma\sim 2$ in GRS 1915+105 (Zdziarski et al.\ 2001).

Finally, we caution that a commonly-used implementation of the BMC
model, {\sc bmc} in the X-ray data fitting code {\sc xspec}, allows
for an arbitrary normalization of the Comptonization tail, arbitrary
value of the spectral index as well as it does not include any
high-energy cutoff, in contrast to the theoretical results. Thus,
satisfactory results of fitting X-ray data with this model cannot be
automatically taken as showing a consistency between that model and
the data.

\section*{Acknowledgements}

We thank L. Titarchuk and Ph.\ Laurent for discussions, and Ph.\
Laurent for supplying his Monte Carlo results. This paper was
supported through KBN grants 1P03D01827, 1P03D01128, 4T12E04727 and
PBZ-KBN-054/P03/2001.

\label{lastpage}

\begin{thebibliography}{}

\bibitem{b72}
Bardeen J.~M., Press W. H., Teukolsky S. A., 1972, ApJ, 178, 347

\bibitem{b79} 
Begelman M.~C., 1979, MNRAS, 187, 237

\bibitem{bp81a} 
Blandford R.~D., Payne D.~G., 1981a, MNRAS, 194, 1033

\bibitem{bp81b} 
Blandford R.~D., Payne D.~G., 1981b, MNRAS, 194, 1041

\bibitem{bg71}
Blumenthal G.~R., Gould R.~J., 1970, RvMP, 42, 237 

\bibitem{b99} 
Borozdin, K., et al. 1999, ApJ, 517, 367

\bibitem{ct95} 
Chakrabarti S. N., Titarchuk L., 1995, ApJ, 455, 623

\bibitem{c83}
Chandrasekhar, S., 1983, The Mathematical Theory of Black Holes. 
Oxford Univ.\ Press, Oxford

\bibitem{c88}
Colpi M., 1988, ApJ, 326, 223

\bibitem{c75}
Cunningham C. T., 1975, ApJ, 202, 788

\bibitem{damico}
D'Amico F., Heindl W.~A., Rothschild R.~E., Gruber, D.~E., 2001, ApJ, 547, L147

\bibitem{disalvo00} 
Di Salvo T., et al., 2000, ApJ, 544, L119

\bibitem{disalvo01} 
Di Salvo T., Robba N.~R., Iaria R., Stella L., Burderi L., Israel G.~L., 2001, ApJ, 
554, 49 

\bibitem{disalvo02} 
Di Salvo T., et al., 2002, A\&A, 386, 535

\bibitem{etc96} 
Ebisawa K., Titarchuk L., Chakrabarti S., 1996, PASJ, 48, 59

\bibitem{f05}
Farinelli R., Frontera F., Zdziarski A.~A., Stella L., Zhang S.~N., van der Klis M., 
Masetti N., Amati L., 2005, A\&A, 434, 25 

\bibitem{gd04}
Gierli{\' n}ski M., Done C., 2004, MNRAS, 347, 885 

\bibitem{g99}
Gierli\'nski M., Zdziarski A. A., Poutanen J., Coppi P. S., Ebisawa K., Johnson
N. W., 1999, MNRAS, 309, 496

\bibitem{gw84}
G\'orecki A., Wilczewski W., 1984, Acta Astronomica, 34, 141

\bibitem{g98}
Grove J.~E., Johnson W.~N., Kroeger R.~A., McNaron-Brown K., Skibo J.~G., 
Phlips B.~F., 1998, ApJ, 500, 899

\bibitem{iaria}
Iaria R., Burderi L., Di Salvo T., La Barbera A., Robba N.~R., 2001, ApJ,  547, 412 

\bibitem{j01} 
Jaroszy\'nski M., 2001, Acta Astronomica, 51, 91

\bibitem{lt99} 
Laurent P., Titarchuk L., 1999, ApJ, 511, 289 (LT99)

\bibitem{lt01} 
Laurent P., Titarchuk L., 2001, ApJ, 562, L67

\bibitem{mz99}
Magdziarz P., Zdziarski A. A., 1995, MNRAS, 273, 837

\bibitem{mc02} 
McConnell M. L., et al., 2002, ApJ, 572, 984

\bibitem{mf03}
Merloni A., Fabian A.~C., 2003, MNRAS, 342, 951 

\bibitem{m02} 
Miller J. M., et al., 2002 ApJ, 570, L69

\bibitem{mc}
Muchotrzeb-Czerny B., 1986, Acta Astronomica, 36, 1 
 
\bibitem{n05} 
Nied\'zwiecki A., 2005, MNRAS, 356, 913 (N05)

\bibitem{pt74} 
Page D. N., Thorne K. S., 1974, ApJ, 191, 499

\bibitem{pp01}
Papathanassiou H., Psaltis D., 2001, MNRAS, submitted (astro-ph/0011447)

\bibitem{pb81}
Payne D.~G., Blandford R.~D., 1981, MNRAS, 196, 781

\bibitem{p77} 
Piran T., Shaham J., 1977, Physical Review D, 16, 1615

\bibitem{pss}
Pozdnyakov L. A., Sobol' I. M., Sunyaev R. A., 1983, Ap.\ Space Phys.\ 
Rev., 2, 189

\bibitem{p01} 
Psaltis D., 2001, ApJ, 555, 786

\bibitem{pl97} 
Psaltis D., Lamb F.~K. 1997, ApJ, 488, 881

\bibitem{rb97}
Reynolds C.~S., Begelman M.~C., 1997, ApJ, 488, 109 

\bibitem{rl79} 
Rybicki G.~B., Lightman A.~L., 1979, Radiative Processes in Astrophysics 
(New York: Wiley)

\bibitem{st98} 
Shrader C., Titarchuk L., 1998, ApJ, 499, L31

\bibitem{st99}
Shrader C., Titarchuk L., 1999, ApJ, 521, L121

\bibitem{tmk97}
Titarchuk L., Mastichiadis A., Kylafis N.~D., 1997, ApJ, 487, 834

\bibitem{tz98}
Titarchuk L., Zannias T., 1998, ApJ, 493, 863	

\bibitem{t99}
Tomsick J. A., Kaaret P., Kroeger R. A., Remillard R. A., 1999,
ApJ, 512, 892

\bibitem{tzt02}
Turolla R., Zane S., Titarchuk L., 2002, ApJ, 576, 349

\bibitem{u02}
Ueda Y., et al., 2002, ApJ, 571, 918

\bibitem{weaver}
Weaver T.~A., 1976, PhRvA, 13, 1563 

\bibitem{aaz00}
Zdziarski A. A., 2000, in P. C. H. Martens, S. Tsuruta \& M. A. Weber, eds., 
IAU Symp.\ 195, Highly Energetic Physical Processes and Mechanisms for Emission 
from Astrophysical Plasmas. ASP, San Francisco, p.\ 153 (astro-ph/0001078)

\bibitem{zg04}
Zdziarski A. A., Gierli\'nski M., 2004, Progr.\ Theor.\ Phys.\ Suppl., 155, 99

\bibitem{aaz01}
Zdziarski A. A., Grove J. E., Poutanen J., Rao A. R., Vadawale S. V., 2001, ApJ, 
554, L45

\bibitem{zcc97}
Zhang S.~N., Cui W., Chen W., 1997, ApJ, 482, L155



\end{thebibliography}
\end{document}